\documentclass[letterpaper]{article}

\usepackage{aaai21}  
\usepackage{times}  
\usepackage{helvet} 
\usepackage{courier}  
\usepackage[hyphens]{url}  
\usepackage{graphicx} 
\usepackage{natbib}  
\usepackage{caption} 
\usepackage{color}
\usepackage{courier}
\usepackage{graphicx}
\usepackage{float}
\usepackage{multirow}
\usepackage{adjustbox}
\usepackage{caption}
\usepackage{amsmath}
\usepackage[dvipsnames]{xcolor}
\usepackage[caption=false]{subfig}
\usepackage{booktabs}

\newcommand{\citenoun}[1]{{\citeauthor{#1} \shortcite{#1}}}
\newcommand{\cut}[1]{}

\makeatletter
\def\thickhline{%
  \noalign{\ifnum0=`}\fi\hrule \@height \thickarrayrulewidth \futurelet
   \reserved@a\@xthickhline}
\def\@xthickhline{\ifx\reserved@a\thickhline
               \vskip\doublerulesep
               \vskip-\thickarrayrulewidth
             \fi
      \ifnum0=`{\fi}}
\makeatother

\newlength{\thickarrayrulewidth}
\begin{document}
\title{Leaders or Followers? A Temporal Analysis of Tweets from IRA Trolls}
\author{Siva  K. Balasubramanian\textsuperscript{\rm 1}, Mustafa Bilgic\textsuperscript{\rm 2}, Aron Culotta\textsuperscript{\rm 3}, Libby Hemphill\textsuperscript{\rm 4}, Anita Nikolich\textsuperscript{\rm 5}, Matthew A. Shapiro\textsuperscript{\rm 6}\\
}
\affiliations{
    \textsuperscript{\rm 1}Stuart School of Business, Illinois Institute of Technology, Chicago, IL\\
    \textsuperscript{\rm 2}Department of Computer Science, Illinois Institute of Technology, Chicago, IL\\
    \textsuperscript{\rm 3}Department of Computer Science, Tulane University, New Orleans, LA\\
    \textsuperscript{\rm 4}School of Information, University of Michigan, Ann Arbor, MI\\
    \textsuperscript{\rm 5}School of Information Sciences, University of Illinois, Urbana-Champaign, IL\\    
    \textsuperscript{\rm 6}Department of Social Sciences, Illinois Institute of Technology, Chicago, IL\\    
}

\maketitle
\begin{abstract}
\begin{quote}
The Internet Research Agency (IRA) influences online political conversations in the United States, exacerbating existing partisan divides and sowing discord. In this paper we investigate the IRA's communication strategies by analyzing trending terms on Twitter to identify cases in which the IRA leads or follows other users.
Our analysis focuses on over 38M tweets posted between 2016 and 2017 from IRA users (n=3,613), journalists (n=976), members of Congress (n=526), and politically engaged users from the general public (n=71,128). We find that the IRA tends to lead on topics related to the 2016 election, race, and entertainment, suggesting that these are areas both of strategic importance as well having the highest potential impact. Furthermore, we identify topics where the IRA has been relatively ineffective, such as tweets on military, political scandals, and violent attacks. Despite many tweets on these topics, the IRA rarely leads the conversation and thus has little opportunity to influence it. We offer our proposed methodology as a way to track the strategic choices of future influence operations in real-time.
\end{quote}
\end{abstract}

\section{Introduction}
\label{sec:introduction}

The efforts of the Internet Research Agency (IRA) to influence online political discourse in the United States in the 2016 presidential election and beyond are by now well-documented \cite{aral2019,mckay2020,lukito2020b}. Especially since Twitter released lists of IRA-associated accounts and tweets, numerous studies have characterized the content of IRA tweets and their retweet networks, even identifying instances where mainstream news sources refer to IRA accounts directly~\cite{lukito2020b}. While compelling, these direct measures of influence are rare, limited in scope, and do not address the potential widespread influence of the IRA's campaigns over political discourse through more subtle means, such as by exacerbating existing partisan divides and sowing discord. While these indirect paths of influence are inherently more difficult to quantify, identifying them could help us better understand the strategies and breadth of such campaigns.

\begin{figure*}
\centering
\subfloat[IRA leading Users on the term ``march,'' referring to the Women's March of January 2017.]{%
\resizebox*{8cm}{!}{\includegraphics{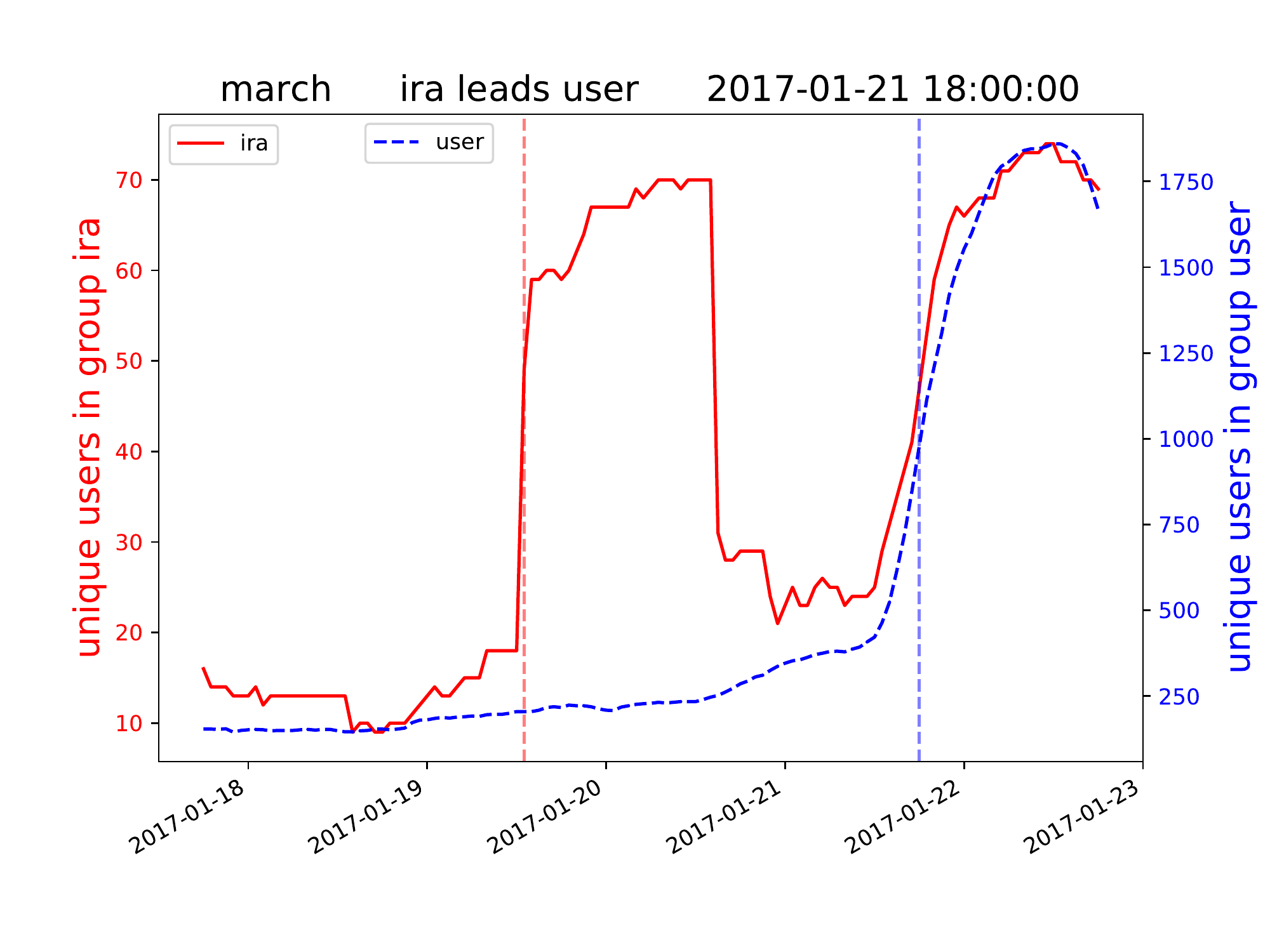}}} \:\:\:\:\:\:\:\:
\subfloat[Users leading the IRA on the term ``health,'' referring to the healthcare debate of October 2016.]{%
\resizebox*{8cm}{!}{\includegraphics{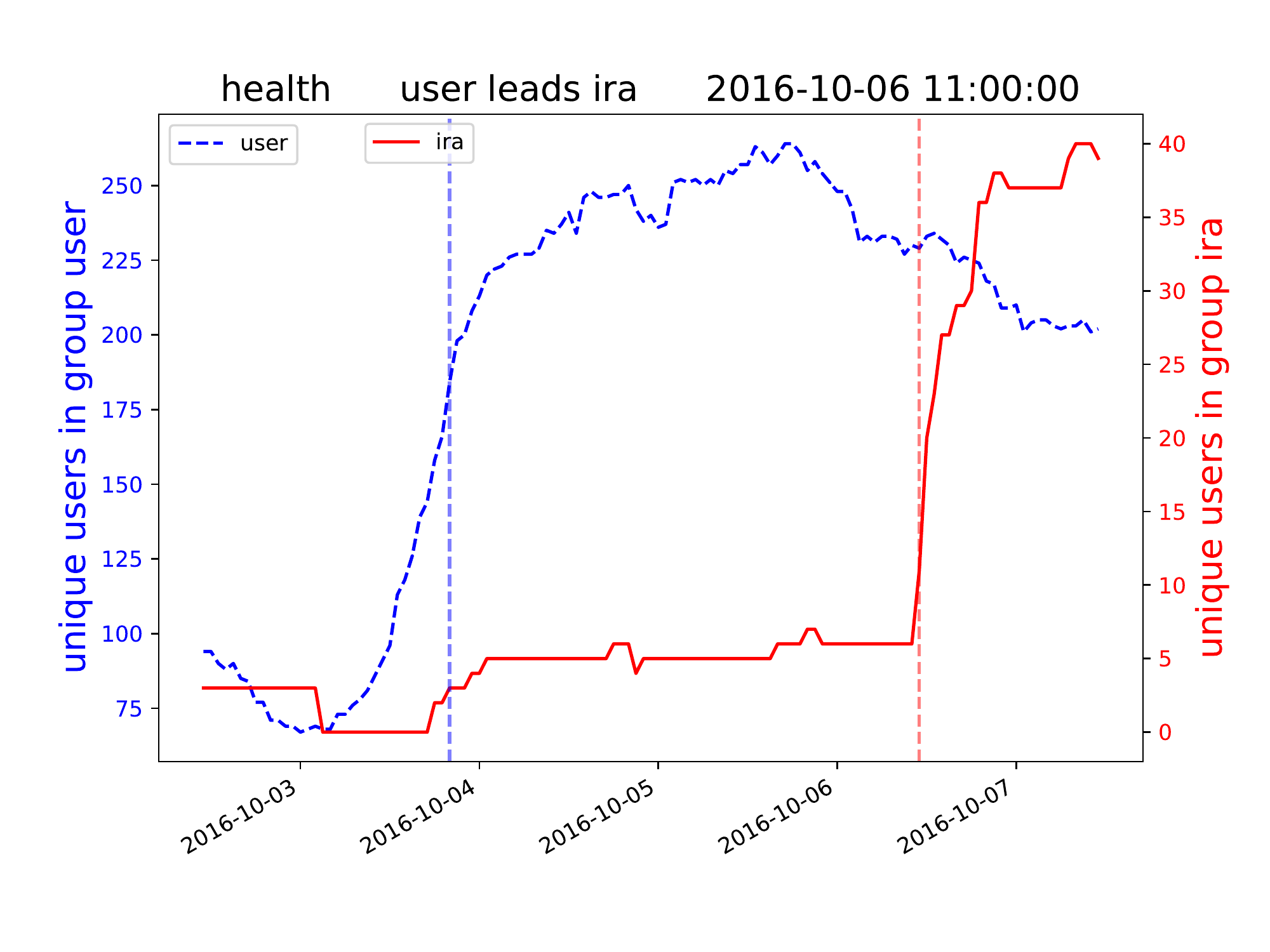}}}
\caption{Two example word spikes, one where the IRA leads and one where it follows the User group.} \label{fig:leading_example}
\end{figure*}

To investigate these issues, in this paper we focus on the \textit{temporal precedence} of salient words on Twitter to distinguish between instances where the IRA is a \textit{leader} or \textit{follower} in a trending conversation. For example, Figure~\ref{fig:leading_example} shows two instances, one where the IRA leads other Twitter users in discussing the Women's March of January 2017, and one where the IRA follows other users in discussing health care in October of 2016. We argue that patterns such as these can provide insight into both the priorities and potential impact of influence campaigns. IRA leadership indicates a greater level of effort to be ``ahead of the curve'' and suggests a greater potential to frame and influence the conversation than messages posted after a trend has already been established. 

The primary contributions of this paper are (1) to establish a methodology for identifying the \textit{temporal precedence} of salient word trends in online media, and (2) to apply it to understand the strategic choices and potential impact of IRA campaigns. We analyze over 38M tweets from the IRA, members of Congress (MOCs), journalists, and ``ordinary'' users posted between January 2016 and December 2017 and investigate the following research questions:

\begin{itemize}
\item \textbf{RQ1: For which topics is the IRA more likely to lead than follow the trend?} We conduct topic clustering analysis to group terms into topics and compute statistics over leading and following frequencies. We find that the IRA is most likely to lead on topics related to the election, race, and entertainment; it is more likely to follow on topics related to the military, health policy, and violent attacks.
\item \textbf{RQ2: How does this temporal precedence vary by user group?} We conduct additional analyses for Twitter users who are MOCs and journalists to assess how the leading and following relationships vary. We find, for example, that journalists tend to lead on topics related to scandals and technology, while MOCs tend to lead on topics related to gun policy and the military.
\item \textbf{RQ3: What is the potential return on investment to the IRA from their efforts to influence conversations on each topic?} By comparing the total number of tweets the IRA posts on a topic with the number of users who tweet after the IRA on that same topic, we can identify the IRA's differential potential impact across topics. For example, even though the IRA tweets with greater frequency about violent attacks and military relative to other topics, they rarely lead the conversations, and so have little opportunity to influence them.
\end{itemize}

In the remainder of this article, we first review some of the extensive background on influence operations in general and the IRA specifically (\S\ref{sec:background}); we then describe our data collection and analysis methods (\S\ref{sec:methods}). Next, we describe the core results (\S\ref{sec:results}) and discuss their implications for protecting online discourse in the future (\S\ref{sec:discussion}), and end with a concluding summary (\S\ref{sec:conclusion}). Additional sensitivity analyses and sample data are included in the Appendix.

\section{Russian Influence Operations and the IRA}
In light of the role the IRA played during the 2016 U.S. presidential election, we contextualize our analysis of the IRA’s strategies and potential influence with a brief overview of Russia’s Influence Operations -- sometimes called Information Operations (IO), other times called Disinformation \cite{jowett2018}. Russia utilized these techniques during the Soviet era for both adversaries and its own citizens, ingraining messages of patriotism and loyalty \cite{bittman1985},\footnote{This is not to say that foreign governments besides Russia do not also engage in these behaviors or that countries besides the U.S. are not attacked. See \citenoun{martin2019} for a broad assessment of foreign influence efforts from 2013 to 2018.} and it has honed its IOs since then to the point where they are a crucial tool of statecraft.\footnote{Rooted in psychology, communications, public relations and operations research, Russia’s weaponized IO originated under Joseph Stalin in 1923, when the KGB’s precursor, the GPU, created a special disinformation office to conduct active intelligence operations. These Soviet operations were under more scrutiny during the Cold War under President Reagan, and were monitored by the U.S. State Department, which put out annual reports such as “Soviet Active Measures: Forgery, Disinformation, Political Operations” until the end of the 1980’s \cite{manning2004}.} Information Warfare (IW), a subset of IO used during the Cold War, further incorporates electronic means of active manipulation \cite{harknett1996}, protecting Russian ideology from Western influence and weaponizing information to influence opinion and foster unique narratives. 
\label{sec:background}

The IRA was formed shortly after the 2012 election of Vladimir Putin, which spurred a slew of domestic Internet censorship measures. Specifically, the IRA focused on bolstering positive sentiment among the populace by hiring workers to write positive content on Russian blogs and sites. While the IRA was officially registered in 2013 to Russian billionaire Yevgeny Prigozhin, who was indicted by the US in 2018 for interfering in the 2016 elections,\footnote{Source: https://www.nytimes.com/2018/02/16/world/europe/prigozhin-russia-indictment-mueller.html.} all evidence indicates that the IRA operates under the direction and authority of the Kremlin. Yet, little was known about the structure of IRA’s operations, tactics, and political goals until Russian undercover reporter Alexandra Garmazhapova published an exposé in 2013.\footnote{Source: https://novayagazeta.ru/articles/2013/09/09/56265-gde-zhivut-trolli-kak-rabotayut-internet-provokatory-v-sankt-peterburge-i-kto-imi-zapravlyaet} For the American audience, the first of two \textit{The New York Times} exposés were published in 2015, drawing attention to the IRA’s existence and ``troll factory'' tactics and highlighting the IRA’s geopolitical goals of sowing discord in countries targeted by the Kremlin as enemies of Russian ideology.\footnote{Sources: https://www.nytimes.com/2015/06/07/magazine/the-agency.html and https://www.nytimes.com/2018/02/18/world/europe/russia-troll-factory.html} 

IRA trolls engage in tasks that are specific but, when coordinated, function much like an industrial effort \cite{linvill2020}, playing ideologies off of each other and working both sides of an issue \cite{golovchenko2020,linvill2020,zhang2021}. Their efforts to organize protests on opposite sides of an issue have had significant consequences, such as motivating African Americans to boycott elections, increasing distrust of political institutions among Latinos, prompting right-wing voters to be confrontational \cite{im2020}, and spreading fake news \cite{howard2019}. Linguistic innovations are often employed by the IRA: e.g., in response to violence involving immigrants, one Russian account tweeted, ``Between the \#rapefugees and the \#refujihadis I think we've all had quite enough of this ‘refugee’ farce.'' The introduction of new linguistic terms like these (i.e. \#rapefugees and \#refujihadis) allows the IRA to frame the debate over immigration as one of national security and violence in order to influence citizens’ reactions and political views.

The IRA also amplifies conversations and messages about particular policy issue areas that may be subsequently read by others, namely MOCs, journalists, and the public. Yet, the IRA approaches ``disinformation in ongoing topics differently based on the political affiliation of their target audience: US conservative audiences are... targeted... about general topics, [while] African American audiences are... targeted with tweets about...Black Lives Matter[, the purpose of which is] to manipulate and radicalize, with some gaining meaningful influence in online communities after months of behavior designed to blend their activities with those of authentic and highly engaged US users'' \cite [27] {howard2019}.

The Senate Select Committee on Intelligence analyzed 10.4M tweets provided by social media platforms in 2017 and found that the IRA’s behavior on Twitter was arguably ``organic.'' That is, topics are chosen in a reactive manner along the lines of a ``digital marketing agency,'' focusing on current events rather than the more careful cultivation of themes on Facebook or Instagram  \cite{diresta2018}. We invoke these claims of ``organic'' Twitter use given the fluid and exploratory nature of the IRA’s activities. For example, trolls employ multiple personas to assess which ones have the greatest impact.\footnote{See \citenoun{xia2019} for a case study of a single IRA persona.} As well, tests for the efficacy of IRA’s statements across different social media platforms suggest that IRA activities on Twitter are often preceded by related Reddit-based activity one week prior 
\cite{lukito2020a}, implying that the IRA was testing its strategies on one platform before using them in a more widespread fashion on other platforms. \citenoun{golovchenko2020} classify this sort of behavior as a ``pre-propaganda strategy.''

The release of historical IRA Twitter datasets in the past few years has led to some new insights into how the IRA's messaging propagated in social media. For example, \citenoun{zannettou2020characterizing} study how images flow from Twitter to Reddit and other platforms; \citenoun{stewart2018examining} examine IRA retweet networks around the \#BlackLivesMatter movement; and \citenoun{badawy2018analyzing} study how retweeting IRA accounts varies by political stance.

Despite the IRA’s continued efforts to affect Americans' exposure to specific narratives \cite{linvill2019}, including recent misinformation posted by the IRA about the efficacy of COVID-19 vaccines \cite{walter2020}, we acknowledge that the IRA may effect little change for Twitter users with extreme political beliefs and attitudes \cite{bail2020,lazer2020}. Yet, the combination of strategies employed by the IRA can serve to bolster messages of dubious accuracy. With sufficient exposure to these types of messages, people begin to treat such information as being more reliable \cite{lewandowsky2012,berinsky2017}, particularly if the information or rumor is discussed at length within a particular social cluster \cite{difonzo2013}.

Given that the effects of general misinformation (i.e., not solely IRA-related misinformation) linger long after one’s exposure to it and even after one has been exposed to corrections \cite{wittenberg2020,nyhan2015,nyhan2010}, attempts to understand how the IRA may have influenced the American public must be rooted in an attempt like ours: to understand how the IRA initially chooses how and what to communicate.\footnote{The steps to achieve this understanding are outlined by \citenoun{aral2019} as follows: identify impressions of manipulative content, match one's impressions to one's voting patterns, establish causality between the two, and then identify the impacts on election outcomes.}

\section{Methodology}
\label{sec:methods}

In this section, we describe our computational approach to data collection, term extraction, time series processing, and topic discovery.\footnote{Replication materials are at: https://github.com/tapilab/icwsm-2022-leader }

\subsection{Data}

\begin{itemize}
\item \textbf{Russian troll accounts (IRA):} We downloaded the October 2018 Twitter release of $\sim$2.9M tweets from 1,635 accounts found to be affiliated with Russia’s Internet Research Agency. 
\item \textbf{Members of Congress accounts (MOCs):} $\sim$2.5M tweets from 526 members of the 115th U.S. Congress (2017-2019) \cite{hemphillschopke2020}. 
\item \textbf{Journalist accounts:} $\sim$900K tweets collected from 976 journalist accounts. These accounts were identified in a semi-automated way. First, we used a query\footnote{The Google query was: ``journalist site:twitter.com inurl:lists''} to identify 20 Twitter Lists that contain the term “journalist,” and then retained accounts that appear on at least two lists to reduce noise. Additionally, we manually searched for the phrases ``liberal journalist'' and ``conservative journalist'' on Twitter’s search page and identified 24 additional accounts. The final list contains a mix of very popular TV personalities (e.g., @AndersonCooper, @WolfBlitzer) as well as many journalists from smaller media outlets across the political spectrum.
\item \textbf{Users:} $\sim$31M tweets from 71,128 ``regular'' users. To identify a sample of politically interested, ordinary Twitter users, we sampled users who follow at least one MOC or journalist collected above. To identify these users, we first collected up to 50K followers of all MOC and journalist accounts (15M unique accounts), and then we sampled 100 followers of each account. We collected up to 5,000 historical tweets from each user, retaining those users who tweeted between 2016-01-01 and 2017-12-31.
\end{itemize}

In total, the dataset contains $\sim$37M tweets from 74,265 users, posted between 2016-01-01 and 2017-12-31.

\subsection{Text processing}

To identify candidate words of interest, we first processed all tokens from IRA accounts by converting them to lowercase; removing punctuation; retaining hashtags, mentions, and emojis; removing URLs; and normalizing numbers. We removed words that occurred in fewer than 50 IRA tweets or more than 40\% of IRA tweets. This resulted in a vocabulary of 47,408 unique terms appearing in $\sim$416M tokens from users from all four groups. To further focus on words that are shared among these four user groups while exhibiting short bursts of popularity, we retained words that are used a minimum of 200 times by at least one of the journalist, MOC, or regular user group while appearing in no more than 1\% of any user group’s tweets. This reduced the vocabulary to 8,535 words. We retained tweets that contained at least one of these words, resulting in $\sim$2.7M IRA tweets, $\sim$2.5M MOC tweets, 880k journalist tweets, and 29.6M user tweets.

\subsection{Time series processing}
\label{sec:tsp}
We define a \textbf{word spike} to be a sharp increase in the usage of a word by a group in a certain time interval, and our overall approach is to identify instances where a word spike for one group immediately precedes a spike in the same word for another group. While such temporal precedence is insufficient to infer a causal relationship between spikes, it provides suggestive evidence regarding the types of words for which IRA is likely to be a \textit{leader} or a \textit{follower}. 

To operationalize this concept, we first construct a time series for each word/user group combination. Let the value $n_{wgt}$ represent the number of unique users from user group $g$ that use the word $w$ in the prior 24 hours from time $t$. We use the number of unique users, rather than tweets, to limit the impacts of a single user tweeting the same term many times. Thus, this value is a measure of the group adoption of a term in the given time period. These values are computed for each hour from 2016-01-01 to 2017-12-31.

We next identify candidate word spike events as follows. For each word time series, we compute the difference vector $\Delta_{wgt} = n_{wgt} - n_{wg(t-1)}$ and retain the top three values, indicating the biggest three “spikes” in the usage of term $w$ by group $g$. To reduce noise, we omit any spikes where $n_{wgt} < 5$. This resulted in 35,826 total word spike events from the four user groups.\footnote{The number of spikes per group is user: 19,940, IRA: 10,066, MOC: 4,282, journalist: 998.}

To identify potential leader-follower relationships between user groups, we select pairs of word spike events $\{ \langle w,g',t' \rangle, \langle w,g,t \rangle \}$ where a sudden increase in the use of word $w$ by group $g$ is immediately preceded by a sudden increase in the use of the same word by another group $g'$. We restrict these pairs to those in which the spike for group $g'$ occurs no more than 4 days prior to the spike for group $g$. This resulted in 10,599 word spike pairs involving 3,415 unique words. 

\begin{figure*}
\centering
\includegraphics[width=\linewidth]{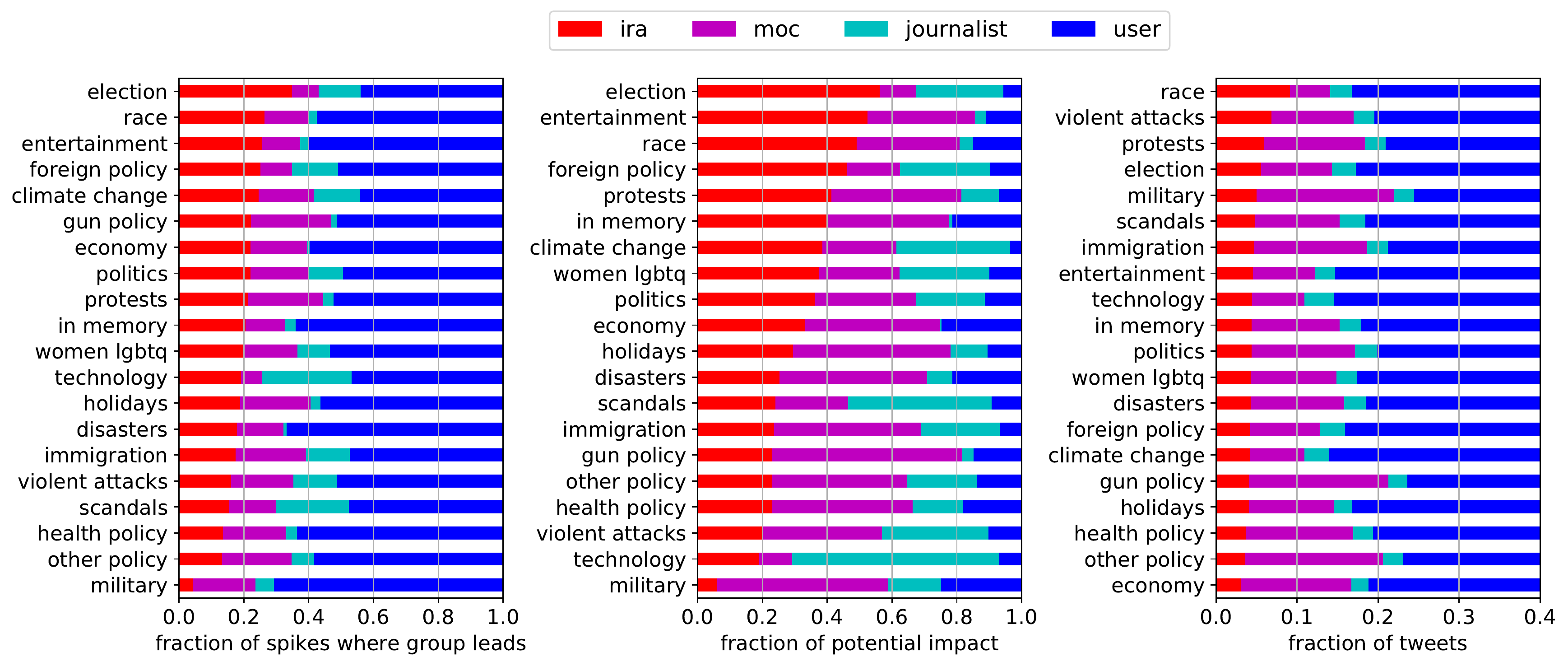}
\caption{\textbf{left panel}: Fraction of word spikes where each group leads; \textbf{center panel}: fraction of total potential impact for each group per cluster; \textbf{right panel}: fraction of all tweets from each group containing at least one term from a cluster } \label{fig:fractions}
\end{figure*}

For each pair of word spike events, we categorize the first group as the \textit{leader} and the second group as the \textit{follower}. As seen previously, Figure~\ref{fig:leading_example} shows two word spike pair events, one in which the IRA leads, and one in which it follows. While the data do not allow us to make causal claims about such events (e.g., we cannot conclusively determine that the leader \textit{caused} the follower to use this term), it is suggestive of a greater level of interest and effort on the part of the leading group with respect to this topic. Furthermore, being a leader on a topic increases the potential for having an influence on that topic. Thus, this methodology allows us to focus on topics for which the IRA has a potential impact, as opposed to examining all topics the IRA discusses while ignoring the evidence of temporal precedence.

\subsection{Term clustering}

To better understand the topics for which the IRA tends to lead other groups, we used a semi-automated approach to cluster terms into meaningful topics. For each of the 3,415 unique words identified in the previous step, we collected all of the corresponding word spike pairs. We then collected the tweets containing each word posted by each user group involved in the leader-follower relationships, restricted to four days prior and one day after the word spike for that group. We then represent each word by a feature vector indicating the count of all other words mentioned in the same tweet as the target word (the context vectors of each word). These context vectors are converted into term frequency-inverse document frequency representations, normalized to unit length.  We then cluster each context vector into one of 500 topics using K-Means clustering.\footnote{The number of clusters was not optimized -- we chose a large number, knowing that human annotators would reduce to a more manageable size in the next step.} For example, one cluster contains the words “\#womensmarch”, “parade,” “inauguration,” in reference to the Women’s March that occurred the day after President Trump’s inauguration in January 2017.

We next manually coded each cluster into topics using an open-ontology approach. Four co-authors independently reviewed each cluster, inspecting the words as well as the contexts in which they appeared, and assigned a label to each cluster. The labels were not pre-specified, but rather were chosen by the annotators separately. These labels were then discussed jointly and merged into a unified schema, resulting in the following 22 topic labels: climate change, disasters, economy, election, entertainment, foreign policy, gun policy, health policy, holidays, immigration, in-memory, military, other, other policy, politics, protests, race, scandals, stopwords, technology, violent attacks, women/LGBTQ. As the “other” and “stopwords” clusters did not contain any semantically meaningful content, we removed them from further analysis, leaving a total of 20 clusters. Table~\ref{tab:clusters} in the Appendix lists the 20 clusters, the number of unique word spike terms in each, and example terms.

\section{Results}
\label{sec:results}

To summarize our research questions from \S\ref{sec:introduction}, we are interested in identifying which topics the IRA leads versus follows (\textbf{RQ1}), how these topics vary by user group (\textbf{RQ2}), and the trade-off between the effort allocated to each topic and its potential for  influencing other conversations (\textbf{RQ3}).

To address \textbf{RQ1}, the left panel of Figure~\ref{fig:fractions} shows the main results by topic, indicating the fraction of word spikes in each topic for which each group leads. For example, we find that of all the term spikes related to the election, IRA leads roughly 35\% of the time. In contrast, the IRA leads less than 5\% of the time for the military topic. Based on this ordering, the IRA appears to lead most often on topics of election, race, and entertainment.
The relatively high ranking of term spikes related to race is consistent with research showing that IRA members are effective when trolling as Black activists \cite{freelon2020}. Yet, given the impact of the IRA on the 2016 U.S. presidential election \cite{aral2019,mckay2020,lukito2020b}, we would have been surprised if the election topic had not been ranked at or at least near the top.

\begin{figure*}
\centering
\includegraphics[width=.9\linewidth]{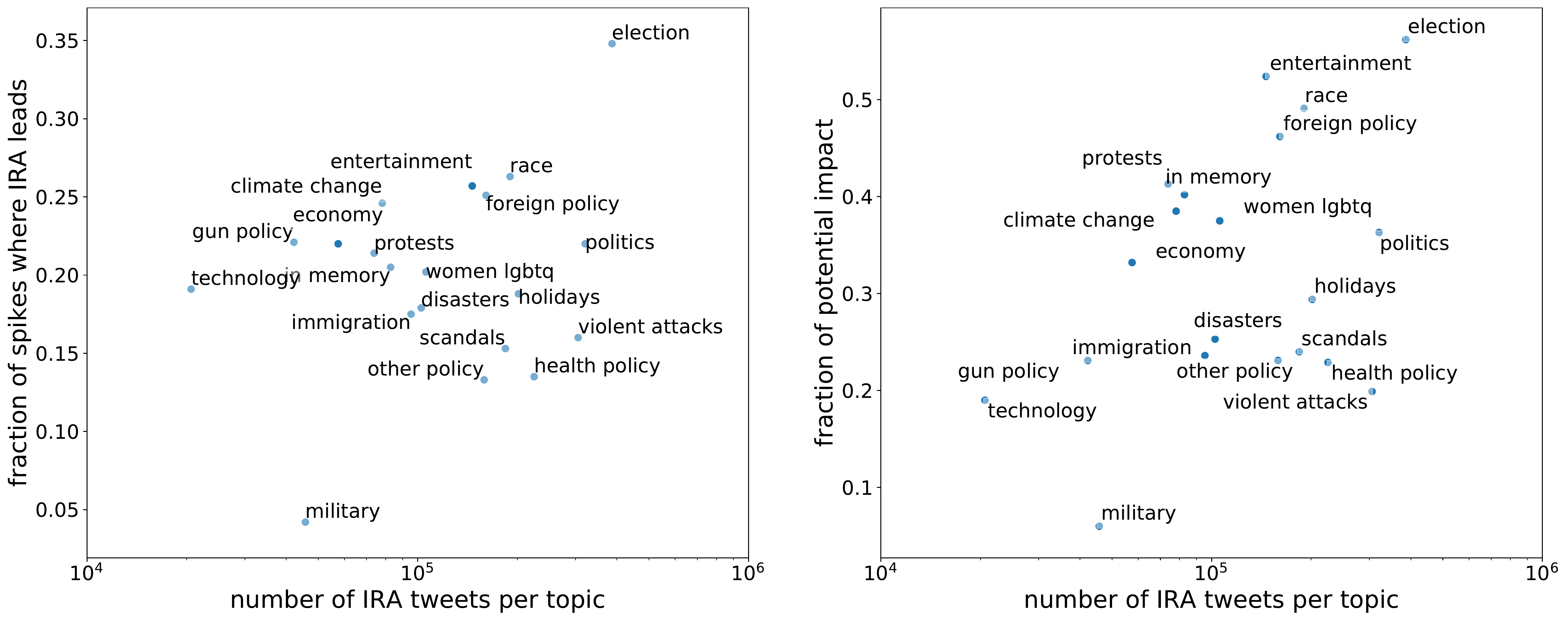}
\caption{Scatter plots showing the relationship between the number of IRA tweets on a topic and the fraction of spikes where the IRA leads \textbf{(left panel)} and the fraction of potential impact \textbf{(right panel)}. Topics in the top left quadrant suggest higher potential return on investment, while topics in the lower right quadrant suggest lower potential returns.  } \label{fig:scatters}
\end{figure*}

This initial analysis, however, ignores the overall volume of each discussion. It could be the case that the IRA leads on many terms in a topic but does not lead on the terms that are involved in the high volume conversations. To address this distinction, we introduce an additional measure called \textit{potential impact}. For a spike pair  $\{ \langle w,g',t' \rangle, \langle w,g,t \rangle \}$, where group $g'$ leads group $g$, the potential impact of group $g'$ is the total number of users in group $g$ who use term $w$ in the four days following $t'$. In other words, it is the total number of users who have the potential to have been influenced by group $g'$, based on temporal precedence. For example, in the ``march'' example from Figure~\ref{fig:leading_example}(a), there are 1,860 users who use the term ``march'' in the four days following the IRA spike on January 19th. We aggregate these values across all spike pairs and group by topic.

The center panel of Figure~\ref{fig:fractions} shows the fraction of potential impact accounted for by each user group per topic. While the ranking of topics is similar to the left panel, there are re-orderings that reflect distinctions between the topics. For example, while the ``protests'' topic is only at rank 9 in the left panel, it rises to rank 5 in the center panel, indicating that, while the IRA does not lead on many conversations about protests, when it does, it leads popular conversations. The converse is true for gun policy, suggesting that, while the IRA often leads such conversations, those conversations are less popular conversations.

Finally, in the right panel of Figure~\ref{fig:fractions}, we report the fraction of tweets from each group that contain at least one word from each topic. Note that the $x$-axis in this figure is truncated since the User group accounts for at least 70\% of all tweets on all topics due to their much greater size. This panel begins to address \textbf{RQ3} -- how does the number of tweets the IRA allocates to a topic relate to the potential impact? For example, while the IRA allocates a significant number of tweets to the ``violent attacks'' topic, both the fraction of spikes where the IRA leads on this topic, as well as the fraction of potential impact, are near the bottom of the list.

This relationship between ``effort'' as measured by the number of tweets on a topic and potential impact and propensity to lead is shown more clearly in Figure~\ref{fig:scatters}. On the $x$-axis is the raw number of tweets from the IRA on each topic ($\log$ scale); on the $y$-axis are the fraction of spikes where the IRA leads and the fraction of potential impact. As implied above, we find the ``violent attacks'' topic in the lower right quadrant of both panels, showing the limited returns on investment in this topic. In contrast, topics like ``entertainment'' and ``protests'' have high potential impact despite the relatively smaller number of tweets on the topic.

The fraction of word spikes for most IRA-led content ranges from 15\% to 25\%. The connection between content and efficiency shown in Figure~\ref{fig:scatters} illustrates how the IRA prioritized the election topic and how, with differentiated efficiency, the IRA utilizes its resources to tweet about other non-election-related topics while focusing on the election. In light of the IRA's limited resources, this multifaceted approach may reflect not just narrative switching, identified in \citenoun{dawson2019}, but also how the IRA engages in an ``organic'' and reactive process, consistent with \citenoun{diresta2018}.

\begin{figure*}
\centering
\includegraphics[width=\linewidth]{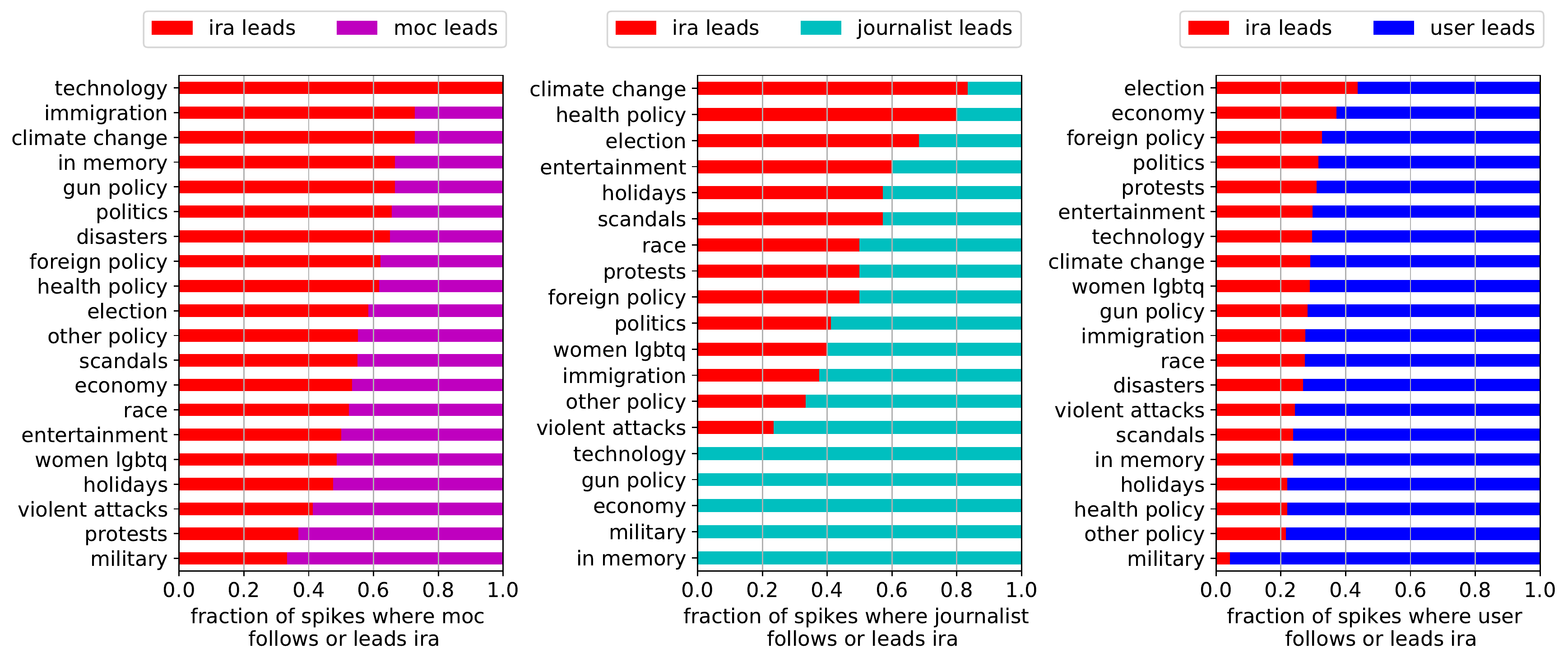}
\caption{\textbf{Left panel}: Of all spike pairs involving IRA and MOC groups, the fraction of spikes for which IRA leads versus follows; \textbf{center panel:} for journalists; \textbf{right panel:} for users.} \label{fig:fractions_by_group}
\end{figure*}

Returning to \textbf{RQ2}, we next look more closely at how the temporal precedence of word spikes varies by user group. To do so, for the MOC, journalist, and user groups, we identify every spike pair involving the IRA group. We then plot by topic the fraction of spikes where the IRA leads or follows the other group. Figure~\ref{fig:fractions_by_group} shows these results for each of the three groups. For example, of the 22 immigration spike pairs involving the IRA and MOC groups, IRA led on 16 of them (73\%), indicated by the immigration row in the first panel. We also observe for the IRA-MOC pair that the IRA leads on most topics, which is consistent with findings related to the 113th Congress showing that MOCs were not likely to lead online discussions about particular issues, tending rather to follow their supporters \cite{barbera2019}. That said, MOCs do lead the IRA on the following topics: military, protests, and violent attacks.   

In terms of the IRA-journalist pair, the center panel of Figure~\ref{fig:fractions_by_group} shows that spikes are predominantly IRA-based with regard to the topics of climate change, health policy, the election, and entertainment. However, the IRA has apparently little interest or is unable to lead the discussion with journalists on the following topics: technology, gun policy, economy, military, and in-memory. The fact that IRA content leads media content across such a large number of topics is potentially problematic given that the media may directly quote IRA tweets \cite{zhang2021,lukito2020b}. Any potential influence by the IRA on journalists would likely reinforce polarization given that information from left and right-leaning media sources is typically consumed by, respectively, people on the left and right \cite{tyler2020}.

Turning now to the IRA-user pair, for no single topic does the IRA lead more than users do. This may reflect the fact that the number of tweets posted by the user group is much greater than those posted by the IRA group. That said, the IRA does lead users on the election topic more than 40\% of the time. The implication of the IRA's temporal precedence on the election topic in particular may have serious effects for conservatives, as they are exposed significantly more to IRA-based information relative to liberals \cite{hjorth2019}, complementing research about 2016 showing that ``top influencers'' who shared news on the left affected Clinton supporters, while Trump supporters affected the behavior of ``top fake news spreaders'' \cite{bovet2019}. 

In addition to the primary empirical results in this section, please refer to the Appendix for a number of sensitivity analyses to assess the robustness of the results to changes in parameter choices, as well as to view additional sample time series to provide further insights into the nature of the leading/following relationships.

\section{Discussion}
\label{sec:discussion}

In terms of election-related content, the data suggest that IRA is clearly focused on being ``ahead of the curve'' -- there is evidence that the IRA's volume of election-related conversations often lead MOCs, journalists, and the general public. The IRA's ability to lead  conversations about elections in social media suggests, but does not conclusively show, their potential to influence them. Future work is needed to more precisely estimate possible causal effects of IRA's efforts on human behavior, especially since unopposed efforts to deliberately mislead can harm American democratic institutions \cite{rodriguez2019,mckay2020}. 

Our results provide insights into how the IRA's temporal precedence in other topics has occurred in the context of content focusing on the election. We acknowledge that the IRA is likely aware that simply targeting political candidates in the context of elections would be an ineffective influence campaign. Thus, the IRA would lead (or follow) non-election topics in an attempt to hide its identity and diversify its image, as discussed in \citenoun{bastos2019}. 

The identification of a rough division of topics between high and low potential impacts and high and low propensities for the IRA to lead is a novel technique. Such impacts and propensities have likely affected the IRA's decisions regarding the design and prioritization of its content. With evidence that posts about certain topics result in disproportionately greater attention, the IRA might redirect its efforts and thus refocus its audiences' attention in order to increase the polarization and radicalization of the American public. The modification of its communication strategy in the wake of receiving this information would further reflect the IRA's application of an ``organic'' and reactive approach \cite{diresta2018}.  

To assess the IRA’s lead-or-follow tendencies, we have distinguished between groups of Twitter accounts that initially present content on certain topics from those that echo and thus reinforce those topics. In terms of what we have accomplished, we characterized the communication strategies among IRA, sometimes leading and sometimes following; we developed quantitative measures to identify words/phrases that are most indicative of each communication strategy; and we identified the temporal priority of a given set of words and topics by either the IRA or one of the other three groups.

Equipped with the techniques and insights offered here, we call for further IRA-focused research in a similar vein. We suggest that research examine the role of exogenous predictors of IRA-based Twitter activity, such as whether fewer word spikes led by the IRA co-occur with Russian holidays or cold weather in St. Petersburg \cite{almond2020}. However, we also encourage future research to consider the presence of second-order (or third-order, etc.) spikes, i.e. instances where a topic initially led by one group is followed up by a second group, but then picked up again by the first group without receiving attribution for having started the discussion in the first place. Through a feedback effect, a second-order event that presents a ``new'' tweet could reinforce an existing narrative or rekindle an old one.

We still do not know definitively whether the benefits of temporal precedence are necessarily greater than those where the IRA may, in times where it does not lead, serve as a conduit to messages posted by others that effectively foster misinformation. The use of multiple dissemination paths would be consistent with the notion of ``rumor cascades'' \cite{friggeri2014}. To address this, in line with \citenoun{weeks2021}, one could examine information flows in the context of network analysis.

\section{Conclusion}
\label{sec:conclusion}

The IRA has influenced political conversations in the United States, 
helping the IRA to foster misinformation, slow down the sharing of accurate information \cite{vosoughi2018}, increase political polarization, all quite likely to undermine deliberative democracy \cite{mckay2020}. While IRA Twitter operations have been described as ``opportunistic real time chatter'' \cite{diresta2018}, the IRA invokes the practice of ``cyber voter interference'' \cite{hansen2019} and continues to modernize and refine the tactics of ``active measures''. These measures are designed to polarize communities and sow doubt about government, a strategy carried over from the Communist era \cite{bittman1985} and adapted for the new media era. 
 
Our broad goal here has been to explain the connections between the content and the temporal precedence of IRA-based Twitter information dissemination to other groups of information receivers, namely MOCs, journalists, and the public. Paths of influence among these four groups are neither uniform, linear, nor simple to predict \cite{zhang2021}. To be explicit, our work is distinct from a growing body of research that address the distinction between rumors, misinformation, and disinformation \cite{guesslyons2020}, the susceptibility of people to misinformation \cite{pennycook2019}, and the means of countering misinformation’s effects through an information campaign of one form or another \cite{wittenberg2020,kuklinski2000}.\footnote{Warning labels may have only modest effects on specific beliefs and virtually none on political parties \cite{guess2020}, but the media itself can reduce public misperceptions with the journalistic fact-checking mechanism \cite{nyhan2020}.} We identified conversations where IRA effectively led, those where it strategically followed, and highlighted potential paths of influence on U.S. politics.

\section{Ethical Considerations}
The data in this paper is derived from publicly-accessible user-generated content online. While our focus is on aggregate trending keywords and not individual user characteristics, such data carry risks for issues of privacy and ``right-to-be-forgotten.'' To mitigate these issues and comply with terms of service, we will release only tweet IDs for the data used in this study.

\bibliography{references}

\newpage
\clearpage

\appendix

\section*{Appendix}
\label{sec:appendix}

\paragraph{Topic list} Table~\ref{tab:clusters} lists the 20 discovered clusters along with the number of unique terms they contain and a sample of the terms.

\paragraph{Bootstrap confidence intervals} To assess the variability of the main results, we constructed a bootstrap distribution using 1,000 samples with replacement of the original set of all word spike pair events, restricted to the top 3 spikes per term/group pair. Tables~\ref{tab:ci_lead} and \ref{tab:ci_impact} report the 95\% confidence intervals for each statistic. The values in Table~\ref{tab:ci_lead} for the fraction of spikes where each group leads generally have narrow intervals, allowing us to have confidence in conclusions, e.g., that the IRA group was more likely to lead on the election topic than others. We do observe a bit more variability in the potential impact values in Table~\ref{tab:ci_impact}, most likely due to the heavy-tailed nature of word spikes. That is, a small number of word spikes involve a very large number of users; removing these from a sample can affect the overall averages. 

\paragraph{Sensitivity analysis by number of spikes} We additionally examine how results vary as we modify the number of spikes considered for each term/group pair. In our main results, we selected the top three spikes for each. Since this number will affect the total number of candidates for leading/following relationships, we investigated how the primary results vary as we consider between 1 and 5 spikes per term/group pair. The results are summarized in Tables~\ref{tab:n_peaks_leads} and \ref{tab:n_peaks_impact}. Using between 2-5 spikes appears to result in very similar results, both in absolute terms and in the ordering of topics. Using a single spike per term/group pair introduces more variablity, most likely due to the smaller sample size.

\paragraph{Sensitivity analysis by spike threshold.} In \S\ref{sec:tsp} we described a filter to omit candidate spikes if fewer than five unique users were involved. Because we retained the top three spikes by difference vector $\delta_{wgt}$, this only filtered a small fraction of spikes. However, given that there are over 70K unique users in our data, we wanted to further check how sensitive the results are to this threshold. We thus removed spikes from the user group if they contained fewer than $N$ users, and varied $N \in [5,10,20,50,100]$. Tables~\ref{tab:size_leads} and \ref{tab:size_impact} shows the results. Generally, the standard deviations across trials are low, between 3-5\%, though increasing the threshold to 100 does appear to influence the magnitude of the results, suggesting that larger samples sizes of users may be needed for additional analysis.

\paragraph{Sensitivity analysis by group size.}  The number of users and tweets from a particular group will affect the nature of the spikes found. In our primary analysis, we attempted to mitigate this by restricting the data to the top three biggest spikes for each group/term pair. E.g., for the term ``march,'' we identified the three biggest spikes for each group for consideration in the leader vs. follower analysis. As an additional check on how this affects results, we performed a sensitivity analysis where we attempt to control for the number of spikes from each group. To do so, we sample 5,000 spikes from the set of all spikes such that each group has the same number of candidate spikes, again drawn from the top three spikes per term/group pair. That is, each group has 1,250 spikes on average. For each such sample, we perform our analysis to determine the leading and potential impact statistics. We repeat this sampling process 1,000 times to generate confidence intervals on the results. Tables~\ref{tab:group_lead} and ~\ref{tab:group_impact} shows the results. 

Generally, the results appear in line with the original results in Table~\ref{tab:ci_lead} and \ref{tab:ci_impact}, though it does appear that the fraction of spikes where the user group leads goes down slightly, whereas the fraction of spikes where the journalist group leads rises slightly. The ordering of topics within each group appears mostly the same --- for example, the highest IRA value remains the election topic, while the lowest remains the military topic.

\begin{figure*}
\centering
\includegraphics[height=9in]{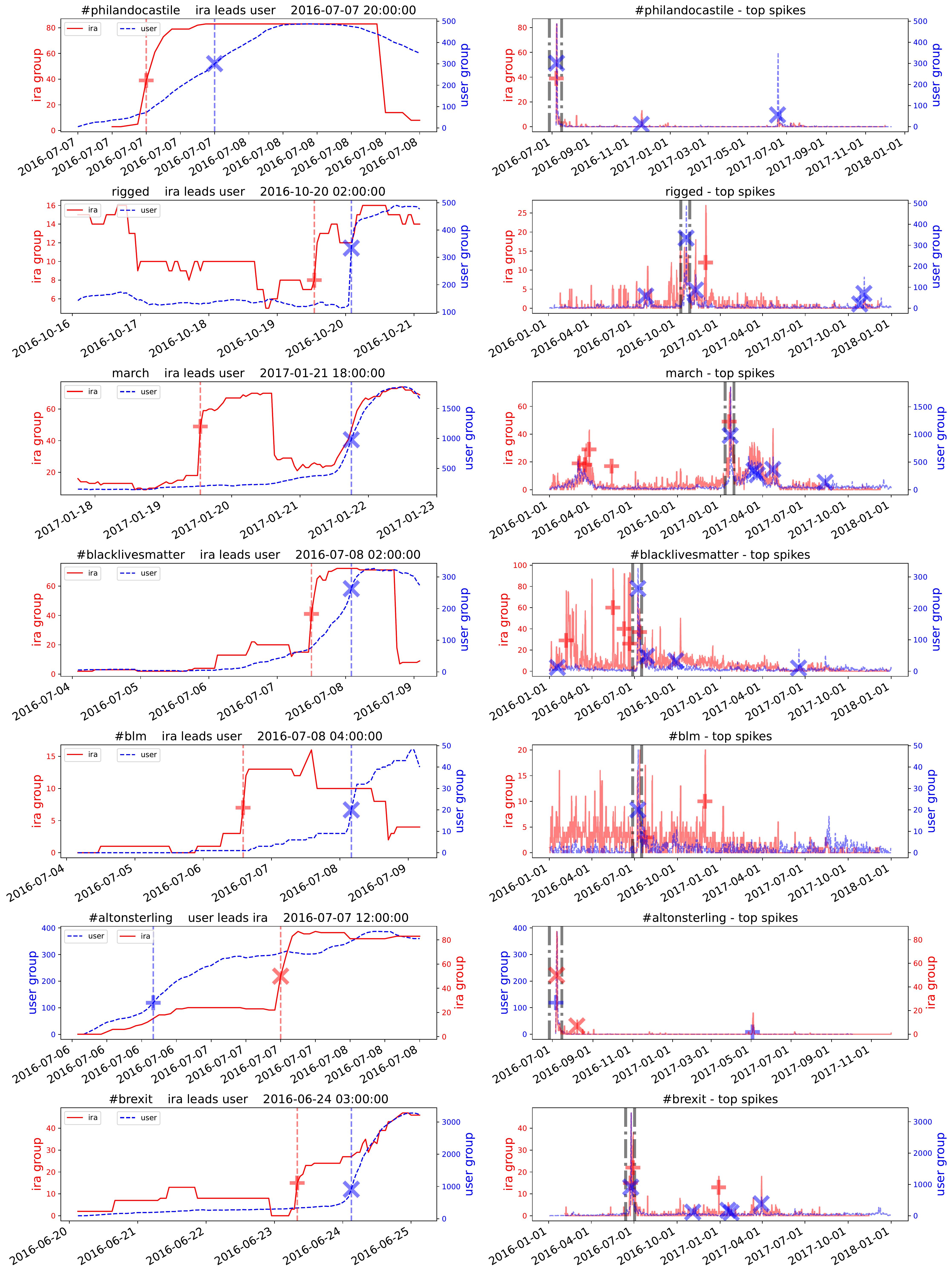}
\caption{Left column: Sample word spike pairs with spike days indicated; Right: complete time series for the corresponding pair on the left, with the top spikes indicated. \label{fig:sample_series}}
\end{figure*}

\begin{figure*}
\centering
\includegraphics[height=9in]{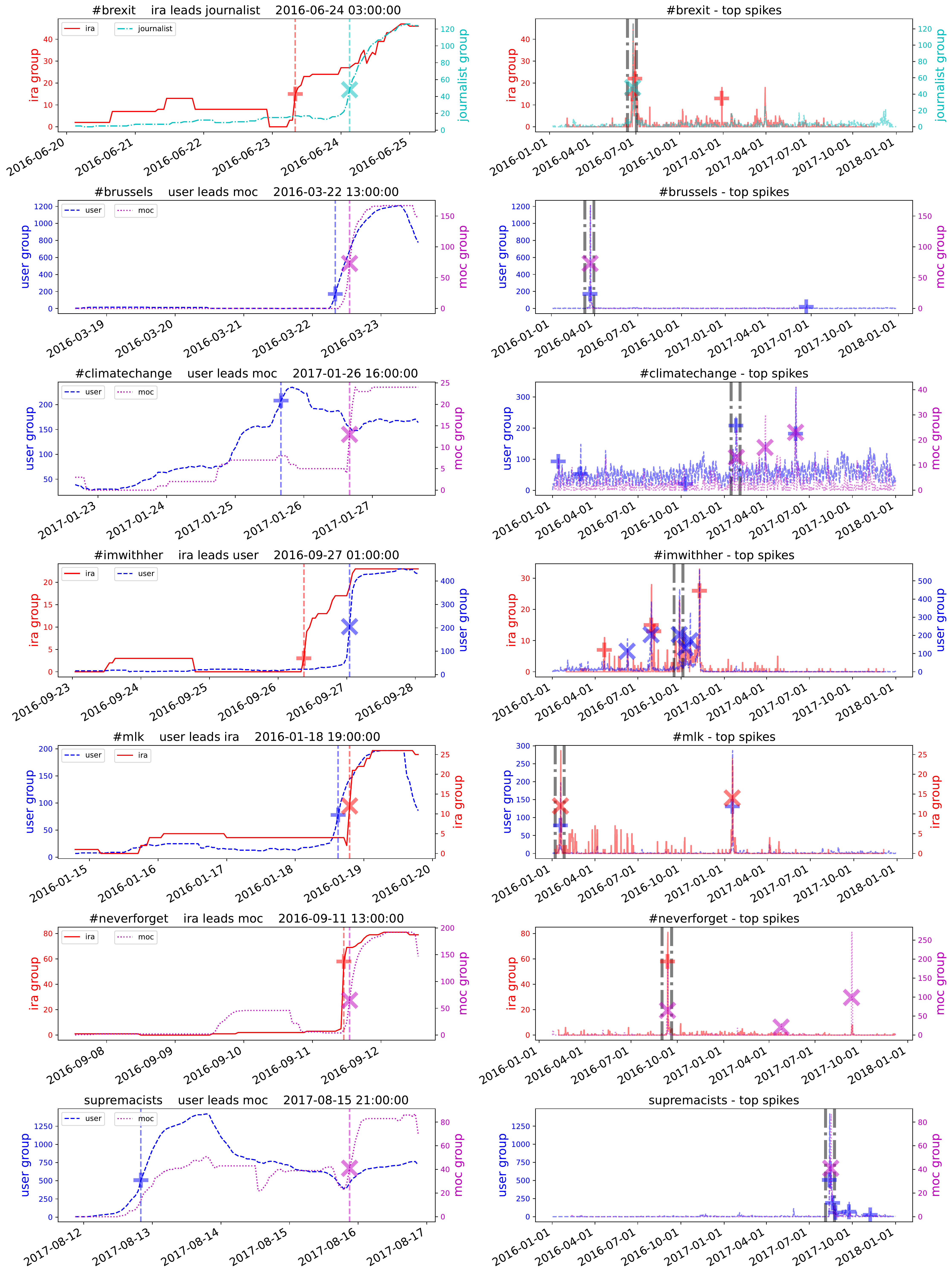}
\caption{Left column: Sample word spike pairs with spike days indicated; Right: complete time series for the corresponding pair on the left, with the top spikes indicated. \label{fig:sample_series2}}
\end{figure*}

\paragraph{Sample time series} To provide some additional insight into the nature of leading and following relationships, Figures \ref{fig:sample_series} and \ref{fig:sample_series2} plot samples of 14 word spike pairs, drawn from the top terms listed in Table~\ref{tab:clusters}. The left columns show the narrow time window for the identified word spikes identified as a leading-follower relationship. The ``+'' and ``X'' symbols indicate the day upon which the spike was identified for the leading and follower groups, respectively. In the right column, the full time series for the term is displayed for the two groups, with up to the top five spikes indicated again by ``+'' and ``X'' symbols. Additionally, the gray vertical bars in the right column highlight the time range shown in more detail by the left column.

Examining these samples reveals a few different types of relationships. For example, for ``march'', ``\#imwithher,'' and ``\#mlk'', the follower group  rarely mentions the term prior to the leading group, after which the number of mentions in the follower group grows rapidly. For other examples, such as ``\#altonsterling,'' ``\#brexit,'' and ``\#climatechange,'' the follower group has a small but growing usage of the term, but a sudden increase in usage only occurs after the leading group's spike. A promising area for future work is to further categorize the nature of these different relationships and to understand their real-world implications on how ideas spread in online networks.
\clearpage

\begin{table}[t]
\small
\centering
\begin{tabular}{p{2.2cm}p{.9cm}p{13.4cm}}
\toprule
\textbf{topic} &  \textbf{unique words} & \textbf{example words}\\
\midrule
climate change &       31 & planet; \#climatemarch; paris; \#climatechange; burning; chinese; temperatures \\
\midrule
disasters &       71 &  hurricane; crisis; puerto; rico; aid; relief; zika; haiti; evacuate; harvey; \#flintwatercrisis; epidemic; opioid \\
\midrule
economy &       38 &   support; economy; small; business;  invest; regulations; unemployment; leaked; owners; employees\\
\midrule
election &      262 &  donald; president; vote;  cia; melania; speech; \#gopdebate; \#inauguration; electoral; @hillaryclinton; giuliani; christie;   \#imwithher; @berniesanders; dnc;  \#trump; \#vote; rigged \\
\midrule
entertainment &      113 &  athletes;  hollywood; worst; \#superbowl; racism; skin; commercial; \#oscars;  \#oscarssowhite; ridiculous\\
\midrule
foreign policy &      103 & minister; national; strike; terror; \#brexit; leave;  castro; missile; syria;  korea;  chemical;  \#brexitornot \\
\midrule
gun policy &       33 & guns; violence; dangerous; background; buying;  demand; legislation; \#holdthefloor;  @nra; bump; stocks \\
\midrule
health policy &      164 & obamacare; health;  planned; parenthood;  coverage; conditions; medicaid; trumpcare; devastating;  mandate\\
\midrule
holidays &      161 &  christmas;  veterans;  \includegraphics[height=1em]{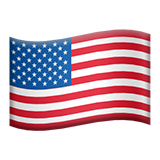}; thanksgiving; valentine's;   memorial; mother's; \#veterans; fourth;  \includegraphics[height=1em]{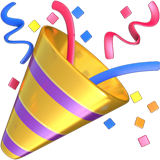};   \includegraphics[height=1em]{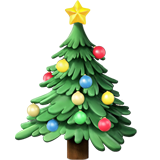};  \includegraphics[height=1em]{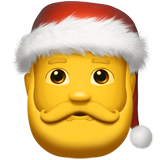}\\
\midrule
immigration &       52 &  ban; immigration; order; refugee; travel;  muslim; refugees; appeals; law; detained; thousands;  unconstitutional; hawaii; latinos; \#refugeeswelcome; \#muslimban;  deportation; undocumented; alien\\
\midrule
in-memory &       55 &   ali;  died; remember;  boxing; muhammad;  rip;  remembering;  \#neverforget;  murdered; pearl; sandy; hook \\
\midrule
 military &       27 &   military; transgender;  qualified; va; serving; marines; whistleblower; accountability; openly\\
\midrule
other policy &       97 &   federal; bill; tax; debt; plan; workers; staff; committee;  corporations;  rich; taxpayers; billionaires; wages\\
\midrule
politics &      220 &                                                 house; press; gorsuch; address; jeff;  justice; nunes;  spicer; scalia;  \#gorsuch; bannon;   floor; legislation\\
\midrule
protests &       37 &    washington; \#womensmarch; science; \#maga; march; signs; \#nodapl; construction; rock; access; oppose; protesters;  petition;  bullets;  pipeline; armed; militia \\
\midrule
race &       58 &  \#mlk; killing;  prison; \#altonsterling; \#blacklivesmatter; tubman;  crime; murder; teen; \#blm; \#philandocastile; cops; sandra; anthem;  \#freddiegray; activist; baltimore; discrimination; disgusting\\
\midrule
scandals &      149 &   sessions; jeff; emails; comey; special; flynn; russia; \#comeyhearing; intel;  sexual; fbi; fired; false; guilty;  classified;  harassment; assault; lying; rigged; obstruction;  firing; tapes; \#comeyday; secrets\\
\midrule
technology &       20 &  rules; iphone; internet; fcc; samsung; pro; apple; protects;  mac; macbook; steal; net\\
\midrule
 violent attacks &      189 &                                   shooting;  isis; attack;  vegas; hate; supremacists; charlottesville; blood; terrorists; \#brussels; barcelona; orlando; mass; lgbt; condolences;  praying; tragedy; blame; unite; horrible;   jewish; threats; heartbreaking\\
 \midrule
 women \& lgbtq &       86 &  \#womensmarch; science; sexual; community; powerful;  military; transgender; lgbt;  gay; harassment; assault;  \#internationalwomensday; rights;  allegations; misconduct; bullying; equality; pledge; openly\\
\bottomrule
\end{tabular}
\caption{The 20 discovered topics with example word spikes.\label{tab:clusters}}
\end{table}

\clearpage

\begin{table*}[t]
\begin{minipage}[t]{\linewidth}\centering
    \footnotesize
    \begin{tabular}{lllll}
    \toprule
    {} &             ira &             moc &      journalist &            user \\
    \midrule
    climate change &  .245 $\pm$ .08 &  .170 $\pm$ .06 &  .145 $\pm$ .07 &  .440 $\pm$ .09 \\
    disasters       &  .179 $\pm$ .06 &  .143 $\pm$ .05 &  .010 $\pm$ .01 &  .668 $\pm$ .07 \\
    economy         &  .220 $\pm$ .08 &  .174 $\pm$ .07 &  .010 $\pm$ .02 &  .596 $\pm$ .09 \\
    election        &  .349 $\pm$ .03 &  .082 $\pm$ .02 &  .130 $\pm$ .02 &  .440 $\pm$ .03 \\
    entertainment   &  .257 $\pm$ .05 &  .115 $\pm$ .04 &  .026 $\pm$ .02 &  .601 $\pm$ .06 \\
    foreign policy  &  .250 $\pm$ .05 &  .098 $\pm$ .03 &  .140 $\pm$ .04 &  .512 $\pm$ .05 \\
    gun policy      &  .220 $\pm$ .08 &  .249 $\pm$ .08 &  .018 $\pm$ .02 &  .512 $\pm$ .09 \\
    health policy   &  .135 $\pm$ .03 &  .196 $\pm$ .04 &  .034 $\pm$ .02 &  .635 $\pm$ .04 \\
    holidays        &  .189 $\pm$ .04 &  .218 $\pm$ .04 &  .031 $\pm$ .02 &  .561 $\pm$ .05 \\
    immigration     &  .175 $\pm$ .05 &  .217 $\pm$ .05 &  .135 $\pm$ .05 &  .474 $\pm$ .07 \\
    in memory       &  .203 $\pm$ .06 &  .123 $\pm$ .05 &  .032 $\pm$ .03 &  .642 $\pm$ .08 \\
    military        &  .043 $\pm$ .05 &  .196 $\pm$ .09 &  .055 $\pm$ .05 &  .706 $\pm$ .10 \\
    other policy    &  .134 $\pm$ .03 &  .214 $\pm$ .04 &  .070 $\pm$ .03 &  .583 $\pm$ .05 \\
    politics        &  .220 $\pm$ .03 &  .181 $\pm$ .03 &  .106 $\pm$ .02 &  .493 $\pm$ .03 \\
    protests        &  .215 $\pm$ .07 &  .231 $\pm$ .07 &  .032 $\pm$ .03 &  .522 $\pm$ .09 \\
    race            &  .263 $\pm$ .06 &  .136 $\pm$ .05 &  .028 $\pm$ .02 &  .574 $\pm$ .07 \\
    scandals        &  .154 $\pm$ .03 &  .144 $\pm$ .03 &  .225 $\pm$ .04 &  .477 $\pm$ .04 \\
    technology      &  .190 $\pm$ .11 &  .066 $\pm$ .07 &  .277 $\pm$ .12 &  .466 $\pm$ .14 \\
    violent attacks &  .161 $\pm$ .03 &  .192 $\pm$ .04 &  .135 $\pm$ .03 &  .511 $\pm$ .04 \\
    women \& lgbtq  &  .203 $\pm$ .05 &  .164 $\pm$ .04 &  .100 $\pm$ .04 &  .533 $\pm$ .06 \\
    \bottomrule
    \end{tabular}
    \caption{Fraction of word spikes where each group leads, with 95\% confidence intervals computed from 1,000 bootstrap samples.\label{tab:ci_lead}}
\end{minipage}
\begin{minipage}[b]{\linewidth}\centering
\footnotesize
\begin{tabular}{lllll}
& & & & \\
\toprule
{} &             ira &             moc &      journalist &            user \\
\midrule
climate change &  .379 $\pm$ .18 &  .232 $\pm$ .13 &  .352 $\pm$ .17 &  .038 $\pm$ .02 \\
disasters       &  .254 $\pm$ .12 &  .451 $\pm$ .15 &  .078 $\pm$ .10 &  .218 $\pm$ .07 \\
economy         &  .334 $\pm$ .16 &  .406 $\pm$ .19 &  .004 $\pm$ .01 &  .256 $\pm$ .11 \\
election        &  .563 $\pm$ .08 &  .112 $\pm$ .05 &  .267 $\pm$ .06 &  .058 $\pm$ .01 \\
entertainment   &  .527 $\pm$ .13 &  .326 $\pm$ .14 &  .036 $\pm$ .04 &  .111 $\pm$ .04 \\
foreign policy  &  .458 $\pm$ .13 &  .166 $\pm$ .09 &  .277 $\pm$ .11 &  .099 $\pm$ .03 \\
gun policy      &  .233 $\pm$ .12 &  .577 $\pm$ .16 &  .038 $\pm$ .06 &  .152 $\pm$ .07 \\
health policy  &  .228 $\pm$ .08 &  .434 $\pm$ .09 &  .154 $\pm$ .09 &  .184 $\pm$ .04 \\
holidays        &  .297 $\pm$ .11 &  .488 $\pm$ .12 &  .108 $\pm$ .10 &  .107 $\pm$ .03 \\
immigration     &  .236 $\pm$ .12 &  .451 $\pm$ .13 &  .246 $\pm$ .12 &  .068 $\pm$ .02 \\
in memory       &  .396 $\pm$ .16 &  .374 $\pm$ .18 &  .012 $\pm$ .01 &  .219 $\pm$ .08 \\
military        &  .064 $\pm$ .09 &  .517 $\pm$ .22 &  .163 $\pm$ .17 &  .256 $\pm$ .13 \\
other policy    &  .233 $\pm$ .10 &  .414 $\pm$ .12 &  .213 $\pm$ .11 &  .140 $\pm$ .04 \\
politics        &  .362 $\pm$ .07 &  .313 $\pm$ .06 &  .211 $\pm$ .06 &  .114 $\pm$ .02 \\
protests        &  .415 $\pm$ .22 &  .401 $\pm$ .22 &  .110 $\pm$ .16 &  .074 $\pm$ .04 \\
race            &  .491 $\pm$ .13 &  .317 $\pm$ .14 &  .041 $\pm$ .05 &  .151 $\pm$ .05 \\
scandals        &  .241 $\pm$ .09 &  .223 $\pm$ .08 &  .442 $\pm$ .10 &  .095 $\pm$ .02 \\
technology      &  .191 $\pm$ .15 &  .106 $\pm$ .13 &  .631 $\pm$ .22 &  .072 $\pm$ .05 \\
violent attacks &  .202 $\pm$ .07 &  .368 $\pm$ .10 &  .327 $\pm$ .10 &  .103 $\pm$ .03 \\
women \& lgbtq  &  .375 $\pm$ .16 &  .248 $\pm$ .14 &  .275 $\pm$ .15 &  .103 $\pm$ .04 \\
\bottomrule
\end{tabular}
\caption{Fraction of total potential impact for each group, with 95\% confidence intervals computed from 1,000 bootstrap samples.\label{tab:ci_impact}}
\end{minipage}
\end{table*}

\clearpage

\begin{table*}
\begin{minipage}[t]{\linewidth}\centering
\footnotesize
\begin{tabular}{lrrrrrr}
\toprule
{} & \multicolumn{5}{c}{number of spikes considered} & \\
{} &     1 &     2 &     3 &     4 &     5 &   std \\
\midrule
climate change  & 0.170 & 0.227 & 0.246 & 0.240 & 0.270 & 0.037 \\
disasters       & 0.134 & 0.161 & 0.179 & 0.183 & 0.187 & 0.022 \\
economy         & 0.200 & 0.226 & 0.220 & 0.200 & 0.188 & 0.016 \\
election        & 0.327 & 0.348 & 0.348 & 0.348 & 0.346 & 0.009 \\
entertainment   & 0.220 & 0.231 & 0.257 & 0.266 & 0.277 & 0.024 \\
foreign policy  & 0.284 & 0.266 & 0.251 & 0.250 & 0.256 & 0.014 \\
gun policy      & 0.178 & 0.197 & 0.221 & 0.233 & 0.226 & 0.023 \\
health policy   & 0.090 & 0.124 & 0.135 & 0.147 & 0.152 & 0.025 \\
holidays        & 0.164 & 0.177 & 0.188 & 0.191 & 0.194 & 0.012 \\
immigration     & 0.161 & 0.186 & 0.175 & 0.162 & 0.154 & 0.013 \\
in memory       & 0.197 & 0.210 & 0.205 & 0.239 & 0.245 & 0.021 \\
military        & 0.071 & 0.038 & 0.042 & 0.066 & 0.067 & 0.016 \\
other policy    & 0.096 & 0.134 & 0.133 & 0.144 & 0.145 & 0.020 \\
politics        & 0.193 & 0.217 & 0.220 & 0.221 & 0.221 & 0.012 \\
protests        & 0.189 & 0.208 & 0.214 & 0.210 & 0.212 & 0.010 \\
race            & 0.310 & 0.264 & 0.263 & 0.261 & 0.264 & 0.021 \\
scandals        & 0.110 & 0.139 & 0.153 & 0.160 & 0.163 & 0.022 \\
technology      & 0.250 & 0.171 & 0.191 & 0.200 & 0.197 & 0.029 \\
violent attacks & 0.132 & 0.147 & 0.160 & 0.170 & 0.175 & 0.017 \\
women \& lgbtq     & 0.188 & 0.198 & 0.202 & 0.209 & 0.214 & 0.010 \\
\bottomrule
\end{tabular}
\caption{Fraction of word spikes where IRA leads, as the number of spikes per word/group pair varies from 1 to 5.\label{tab:n_peaks_leads}}
\end{minipage}
%
\begin{minipage}[t]{\linewidth}\centering
\footnotesize
\begin{tabular}{lrrrrrr}
&&&&&&\\
\toprule
{} & \multicolumn{5}{c}{number of spikes considered} & \\
{} &     1 &     2 &     3 &     4 &     5 &   std \\
\midrule
climate change  & 0.282 & 0.359 & 0.385 & 0.385 & 0.407 & 0.049 \\
disasters       & 0.185 & 0.227 & 0.253 & 0.256 & 0.261 & 0.032 \\
economy         & 0.231 & 0.399 & 0.332 & 0.319 & 0.301 & 0.060 \\
election        & 0.570 & 0.551 & 0.562 & 0.542 & 0.542 & 0.012 \\
entertainment   & 0.491 & 0.479 & 0.524 & 0.508 & 0.535 & 0.023 \\
foreign policy  & 0.580 & 0.528 & 0.462 & 0.463 & 0.474 & 0.051 \\
gun policy      & 0.254 & 0.228 & 0.231 & 0.261 & 0.267 & 0.018 \\
health policy   & 0.188 & 0.240 & 0.229 & 0.241 & 0.252 & 0.025 \\
holidays        & 0.196 & 0.278 & 0.294 & 0.288 & 0.294 & 0.042 \\
immigration     & 0.176 & 0.235 & 0.236 & 0.234 & 0.230 & 0.026 \\
in memory       & 0.359 & 0.416 & 0.402 & 0.440 & 0.440 & 0.034 \\
military        & 0.027 & 0.018 & 0.060 & 0.096 & 0.088 & 0.035 \\
other policy    & 0.230 & 0.247 & 0.231 & 0.228 & 0.222 & 0.009 \\
politics        & 0.316 & 0.361 & 0.363 & 0.358 & 0.349 & 0.020 \\
protests        & 0.384 & 0.415 & 0.413 & 0.411 & 0.412 & 0.013 \\
race            & 0.576 & 0.473 & 0.491 & 0.485 & 0.474 & 0.043 \\
scandals        & 0.199 & 0.235 & 0.240 & 0.256 & 0.251 & 0.022 \\
technology      & 0.145 & 0.144 & 0.190 & 0.210 & 0.232 & 0.039 \\
violent attacks & 0.174 & 0.194 & 0.199 & 0.223 & 0.235 & 0.024 \\
women \& lgbtq     & 0.347 & 0.374 & 0.375 & 0.367 & 0.362 & 0.011 \\
\bottomrule
\end{tabular}
\caption{Fraction of total potential impact for IRA, as the number of spikes per word/group pair varies from 1 to 5.\label{tab:n_peaks_impact}}
\end{minipage}
\end{table*}

\clearpage

\begin{table*}[t]
\begin{minipage}[t]{\linewidth}\centering
\footnotesize
\begin{tabular}{lrrrrrr}
\toprule
{} & \multicolumn{5}{c}{minimum spike threshold} & \\
{} &     5 &    10 &    20 &    50 &   100 &   std \\
\midrule
climate change  & 0.246 & 0.246 & 0.248 & 0.276 & 0.292 & 0.021 \\
disasters       & 0.179 & 0.179 & 0.179 & 0.217 & 0.290 & 0.048 \\
economy         & 0.220 & 0.220 & 0.220 & 0.238 & 0.283 & 0.027 \\
election        & 0.348 & 0.348 & 0.351 & 0.377 & 0.404 & 0.025 \\
entertainment   & 0.257 & 0.257 & 0.259 & 0.300 & 0.418 & 0.070 \\
foreign policy  & 0.251 & 0.251 & 0.252 & 0.274 & 0.328 & 0.033 \\
gun policy      & 0.221 & 0.221 & 0.221 & 0.279 & 0.328 & 0.048 \\
health policy   & 0.135 & 0.135 & 0.138 & 0.164 & 0.217 & 0.035 \\
holidays        & 0.188 & 0.188 & 0.192 & 0.227 & 0.262 & 0.032 \\
immigration     & 0.175 & 0.175 & 0.178 & 0.195 & 0.224 & 0.021 \\
in memory       & 0.205 & 0.205 & 0.209 & 0.222 & 0.283 & 0.033 \\
military        & 0.042 & 0.042 & 0.042 & 0.058 & 0.100 & 0.025 \\
other policy    & 0.133 & 0.133 & 0.134 & 0.154 & 0.207 & 0.032 \\
politics        & 0.220 & 0.220 & 0.223 & 0.252 & 0.302 & 0.035 \\
protests        & 0.214 & 0.214 & 0.218 & 0.239 & 0.308 & 0.040 \\
race            & 0.263 & 0.263 & 0.270 & 0.309 & 0.374 & 0.048 \\
scandals        & 0.153 & 0.154 & 0.155 & 0.168 & 0.189 & 0.015 \\
technology      & 0.191 & 0.191 & 0.196 & 0.225 & 0.231 & 0.019 \\
violent attacks & 0.160 & 0.161 & 0.163 & 0.175 & 0.193 & 0.014 \\
women \& lgbtq     & 0.202 & 0.202 & 0.206 & 0.236 & 0.290 & 0.038 \\
\bottomrule
\end{tabular}
\caption{Fraction of word spikes where IRA leads, as the minimum threshold on spike size for the user group varies.\label{tab:size_leads}}
\end{minipage}
\begin{minipage}[t]{\linewidth}\centering
\footnotesize
\begin{tabular}{lrrrrrr}
&&&&&&\\
\toprule
{} & \multicolumn{5}{c}{minimum spike threshold} & \\
{} &     5 &    10 &    20 &    50 &   100 &   std \\
\midrule
climate change  & 0.385 & 0.385 & 0.385 & 0.387 & 0.388 & 0.002 \\
disasters       & 0.253 & 0.253 & 0.253 & 0.263 & 0.270 & 0.008 \\
economy         & 0.332 & 0.332 & 0.332 & 0.341 & 0.349 & 0.008 \\
election        & 0.562 & 0.562 & 0.562 & 0.566 & 0.568 & 0.003 \\
entertainment   & 0.524 & 0.524 & 0.525 & 0.536 & 0.558 & 0.015 \\
foreign policy  & 0.462 & 0.462 & 0.462 & 0.467 & 0.476 & 0.006 \\
gun policy      & 0.231 & 0.231 & 0.231 & 0.240 & 0.242 & 0.006 \\
health policy   & 0.229 & 0.229 & 0.230 & 0.238 & 0.250 & 0.009 \\
holidays        & 0.294 & 0.294 & 0.294 & 0.299 & 0.302 & 0.004 \\
immigration     & 0.236 & 0.236 & 0.237 & 0.238 & 0.241 & 0.002 \\
in memory       & 0.402 & 0.402 & 0.403 & 0.415 & 0.431 & 0.013 \\
military        & 0.060 & 0.060 & 0.060 & 0.065 & 0.074 & 0.006 \\
other policy    & 0.231 & 0.231 & 0.231 & 0.236 & 0.248 & 0.007 \\
politics        & 0.363 & 0.363 & 0.363 & 0.370 & 0.381 & 0.008 \\
protests        & 0.413 & 0.413 & 0.414 & 0.419 & 0.429 & 0.007 \\
race            & 0.491 & 0.491 & 0.493 & 0.507 & 0.524 & 0.014 \\
scandals        & 0.240 & 0.240 & 0.240 & 0.243 & 0.246 & 0.002 \\
technology      & 0.190 & 0.190 & 0.191 & 0.194 & 0.175 & 0.007 \\
violent attacks & 0.199 & 0.199 & 0.200 & 0.200 & 0.200 & 0.001 \\
women \& lgbtq     & 0.375 & 0.375 & 0.375 & 0.383 & 0.395 & 0.009 \\
\bottomrule
\end{tabular}
\caption{Fraction of total potential impact for IRA, as the minimum threshold on spike size for the user group varies.\label{tab:size_impact}}
\end{minipage}
\end{table*}

\clearpage

\begin{table*}[t]
\begin{minipage}[t]{\linewidth}\centering
\footnotesize
\begin{tabular}{lllll}
\toprule
{} &             ira &             moc &      journalist &            user \\
\midrule
climate change  &  .248 $\pm$ .05 &  .207 $\pm$ .05 &  .256 $\pm$ .03 &  .288 $\pm$ .08 \\
disasters       &  .187 $\pm$ .05 &  .188 $\pm$ .05 &  .022 $\pm$ .01 &  .604 $\pm$ .07 \\
economy         &  .213 $\pm$ .07 &  .236 $\pm$ .07 &  .019 $\pm$ .00 &  .533 $\pm$ .09 \\
election        &  .315 $\pm$ .03 &  .118 $\pm$ .02 &  .255 $\pm$ .01 &  .312 $\pm$ .03 \\
entertainment   &  .249 $\pm$ .06 &  .189 $\pm$ .05 &  .066 $\pm$ .01 &  .496 $\pm$ .07 \\
foreign policy  &  .243 $\pm$ .03 &  .131 $\pm$ .02 &  .255 $\pm$ .02 &  .371 $\pm$ .05 \\
gun policy      &  .252 $\pm$ .06 &  .302 $\pm$ .07 &  .034 $\pm$ .01 &  .412 $\pm$ .08 \\
health policy   &  .150 $\pm$ .02 &  .232 $\pm$ .03 &  .066 $\pm$ .01 &  .551 $\pm$ .04 \\
holidays        &  .204 $\pm$ .03 &  .303 $\pm$ .04 &  .067 $\pm$ .01 &  .426 $\pm$ .05 \\
immigration     &  .176 $\pm$ .04 &  .234 $\pm$ .04 &  .233 $\pm$ .02 &  .356 $\pm$ .05 \\
in memory       &  .225 $\pm$ .06 &  .170 $\pm$ .05 &  .071 $\pm$ .01 &  .533 $\pm$ .08 \\
military        &  .054 $\pm$ .03 &  .238 $\pm$ .08 &  .109 $\pm$ .03 &  .600 $\pm$ .10 \\
other policy    &  .136 $\pm$ .03 &  .262 $\pm$ .04 &  .134 $\pm$ .01 &  .468 $\pm$ .05 \\
politics        &  .224 $\pm$ .02 &  .213 $\pm$ .02 &  .191 $\pm$ .01 &  .372 $\pm$ .03 \\
protests        &  .207 $\pm$ .06 &  .326 $\pm$ .07 &  .064 $\pm$ .01 &  .403 $\pm$ .09 \\
race            &  .279 $\pm$ .07 &  .216 $\pm$ .05 &  .067 $\pm$ .01 &  .439 $\pm$ .08 \\
scandals        &  .144 $\pm$ .02 &  .162 $\pm$ .02 &  .370 $\pm$ .02 &  .324 $\pm$ .03 \\
technology      &  .125 $\pm$ .09 &  .065 $\pm$ .06 &  .527 $\pm$ .11 &  .283 $\pm$ .13 \\
violent attacks &  .142 $\pm$ .02 &  .243 $\pm$ .03 &  .248 $\pm$ .02 &  .367 $\pm$ .03 \\
women \& lgbtq     &  .191 $\pm$ .04 &  .217 $\pm$ .04 &  .192 $\pm$ .02 &  .399 $\pm$ .05 \\
\bottomrule
\end{tabular}
\caption{Fraction of word spikes where each group leads, controlling for group size. 95\% confidence intervals from 1,000 bootstrap samples.\label{tab:group_lead}}
\end{minipage}
\begin{minipage}[t]{\linewidth}\centering
\footnotesize
\begin{tabular}{lllll}
\toprule
{} &             ira &             moc &      journalist &            user \\
\midrule
climate change  &  .191 $\pm$ .14 &  .208 $\pm$ .10 &  .578 $\pm$ .13 &  .023 $\pm$ .01 \\
disasters       &  .160 $\pm$ .10 &  .489 $\pm$ .15 &  .156 $\pm$ .07 &  .195 $\pm$ .07 \\
economy         &  .244 $\pm$ .16 &  .493 $\pm$ .19 &  .009 $\pm$ .00 &  .254 $\pm$ .11 \\
election        &  .330 $\pm$ .07 &  .123 $\pm$ .05 &  .505 $\pm$ .06 &  .042 $\pm$ .01 \\
entertainment   &  .376 $\pm$ .16 &  .445 $\pm$ .15 &  .087 $\pm$ .03 &  .093 $\pm$ .04 \\
foreign policy  &  .267 $\pm$ .11 &  .164 $\pm$ .08 &  .495 $\pm$ .09 &  .075 $\pm$ .02 \\
gun policy      &  .203 $\pm$ .10 &  .593 $\pm$ .13 &  .069 $\pm$ .02 &  .135 $\pm$ .06 \\
health policy   &  .147 $\pm$ .06 &  .426 $\pm$ .08 &  .272 $\pm$ .05 &  .155 $\pm$ .03 \\
holidays        &  .185 $\pm$ .08 &  .514 $\pm$ .10 &  .219 $\pm$ .06 &  .082 $\pm$ .02 \\
immigration     &  .126 $\pm$ .08 &  .416 $\pm$ .10 &  .405 $\pm$ .08 &  .053 $\pm$ .02 \\
in memory       &  .327 $\pm$ .17 &  .441 $\pm$ .19 &  .027 $\pm$ .01 &  .205 $\pm$ .09 \\
military        &  .042 $\pm$ .06 &  .468 $\pm$ .21 &  .283 $\pm$ .12 &  .207 $\pm$ .09 \\
other policy    &  .132 $\pm$ .07 &  .390 $\pm$ .10 &  .371 $\pm$ .07 &  .107 $\pm$ .03 \\
politics        &  .229 $\pm$ .05 &  .309 $\pm$ .06 &  .372 $\pm$ .04 &  .091 $\pm$ .01 \\
protests        &  .243 $\pm$ .20 &  .457 $\pm$ .20 &  .240 $\pm$ .11 &  .060 $\pm$ .03 \\
race            &  .371 $\pm$ .15 &  .418 $\pm$ .14 &  .096 $\pm$ .03 &  .116 $\pm$ .05 \\
scandals        &  .122 $\pm$ .06 &  .182 $\pm$ .05 &  .636 $\pm$ .06 &  .060 $\pm$ .01 \\
technology      &  .081 $\pm$ .07 &  .069 $\pm$ .06 &  .820 $\pm$ .10 &  .029 $\pm$ .02 \\
violent attacks &  .103 $\pm$ .05 &  .322 $\pm$ .07 &  .508 $\pm$ .06 &  .067 $\pm$ .01 \\
women \& lgbtq     &  .200 $\pm$ .12 &  .243 $\pm$ .11 &  .482 $\pm$ .11 &  .075 $\pm$ .02 \\
\bottomrule
\end{tabular}
\caption{Fraction of total potential impact for each group, controlling for group size. 95\% confidence intervals from 1,000 bootstrap samples.\label{tab:group_impact}}
\end{minipage}
\end{table*}

\clearpage

\clearpage

\end{document}